# 1,000-Fold Enhancement of Light-Induced Magnetism in Plasmonic Au Nanoparticles

**Oscar Hsu-Cheng Cheng[1], Dong Hee Son[1,2]\*, Matthew Sheldon[1,3]\***

[1]Department of Chemistry, Texas A&M University, TX, USA
[2]Center for Nanomedicine, Institute for Basic Science, Seoul, Republic of Korea
[3]Department of Material Science & Engineering, Texas A&M University, TX, USA
\*e-mail: dhson@chem.tamu.edu; sheldonm@tamu.edu

**Strategies for ultrafast optical control of magnetism have been a topic of intense research for several decades because of the potential impact in technologies such as magnetic memory[1,2], spintronics[3], and quantum computation, as well as the opportunities for non-linear optical control and modulation[4] in applications such as optical isolation and non-reciprocity[5]. Here we report the first experimental quantification of optically induced magnetization in plasmonic Au nanoparticles due to the inverse Faraday effect (IFE). The induced magnetic moment in nanoparticles is found to be ~1,000x larger than that observed in bulk Au, and ~20x larger than the magnetic moment from optimized magnetic nanoparticle colloids such as magnetite[6]. Furthermore, the magnetization and demagnetization kinetics are instantaneous within the sub-picosecond time resolution of our study, supporting a mechanism of coherent transfer of angular momentum from the circularly polarized excitation to the orbital angular momentum of the electron gas.**

The IFE is an opto-magnetic phenomenon that manifests as an induced magnetization, $\vec{M}_{ind}$, that is parallel or anti-parallel with the axis of circularly polarized excitation based on the helicity of the radiation (Fig. 1b)[7]. This induced magnetization can be described by[8]:

$$\vec{M}_{ind} = \frac{\lambda\upsilon}{2\pi c}(I_{RHCP} - I_{LHCP}) \tag{1}$$

where λ is the wavelength of light in vacuum, υ is the Verdet constant, c is the speed of light, and $I_{RHCP}$ ($I_{LHCP}$) is the incident intensity of right- (left-) handed circularly polarized light. The IFE was first proposed as a reciprocal consequence of the conventional Faraday effect (Fig. 1a)[7], and both effects can be related through the same Verdet constant in equation (2) that describes Faraday rotation:

$$\Delta\theta = \upsilon\vec{B}_{app}L \tag{2}$$

where Δθ is Faraday rotation angle, $\vec{B}_{app}$ is the applied magnetic field, and L is the effective travel distance of light inside the medium.

The IFE has been studied extensively in materials with large Verdet constants, such as $Tb_2Ti_2O_7$ and $Ta_3Ga_5O_{12}$ for application in optically-written magnetic hard drives[9,10]. Notably, strong resonant field concentration from plasmonic metasurfaces interfaced with these materials can further enhance sub-wavelength magnetization[11]. Circularly polarized femtosecond laser pulses can also be used to non-thermally align electron spins in magnetic materials via the IFE[12]. However, there has been limited research studying magnetization that occurs within non-magnetic plasmonic metals due to the IFE, without other magnetic structures, despite compelling theoretical studies that suggest plasmonic nanomaterials may out-perform more conventional magnetic materials in terms of induced magnetic field strength, spatial confinement, ultrafast time response, and other technologically relevant optoelectronic behavior[13-16]. Several studies have also reported anomalously large Verdet constant in plasmonic colloids, that we confirm below, further suggesting strong enhancement of the IFE may also be possible[17,18].

For example, as opposed to the traditional phenomenological analysis that treats the IFE as a nonlinear four-wave mixing process[19,20], Hertel *et al.* recently derived an intuitive microscopic model based on the continuity equation for an electron plasma subject to circularly polarized light[13]. By Hertel's analysis the photo-induced magnetization $\vec{M}_{ind}$ is:

$$\vec{M}_{ind} = \frac{e\varepsilon_0 \omega_p^2}{4m_e\omega^3}\left(i\vec{E}_0 \times \vec{E}_0^*\right) \tag{3}$$

where $m_e$ is electron mass, $\omega$ is frequency of incident light, $\omega_p$ is the plasmon frequency, $\vec{E}_0$ and $\vec{E}_0^*$ are the electric field and complex conjugate of the incident radiation[13,21]. Magnetization is interpreted as resulting from the solenoid-like path traced out by each electron subject to the rotating electric field during an optical cycle. Additionally, our laboratory has recently performed computational analyses of the IFE in plasmonic Au nanoparticles building from Hertel's approach[15]. Our calculations predict pronounced macroscopic drift currents that circulate the exterior of nanoparticles normal to the incident beam, in addition to the coherent solenoid-like motion of all electrons at the optical frequency[22]. Remarkably these associated phenomena of the IFE are enhanced by several orders of magnitude in nanoparticles compared with bulk films, due to resonant optical concentration at plasmonic hot spots[13,15]. Indeed, a recent theoretical study using a quantum hydrodynamic model predicted the generation of magnetic moments up to 0.15 $\mu_B$ per gold atom in 5 nm diameter Au nanoparticles, an enhancement of approximately two orders of magnitude compared with bulk Au films[16]. This mechanistic picture of the light-induced magnetization is also predicted by *ab initio* calculations that show circularly polarized optical fields driving the coherent motion of the electron gas, with magnetization resulting from the induced orbital angular momentum of electrons[23]. Because spin alignment of individual electrons does not contribute to the mechanism, the magnetization is expected to be instantaneously synchronized with an optical pump, in contrast with dynamics that are limited by spin-spin or spin-lattice relaxation times typical in ferromagnetic materials[24,25]. Despite these intriguing predictions, measurements of the magnetic field due to the IFE in plasmonic nanoparticles has not been reported, with no time-resolved studies until now.

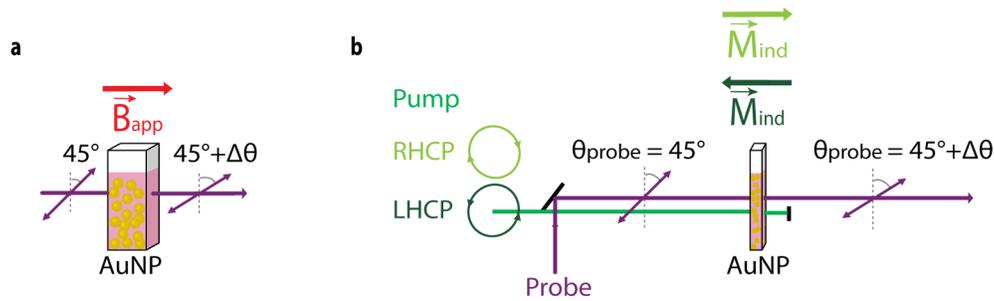

**Figure 1 | Schematic illustration of the Faraday effect and the inverse Faraday effect (IFE) in Au nanoparticle colloids (AuNP). a,** The Faraday effect is the rotation of the polarization plane of light transmitted through a magnetized medium. **b,** The inverse Faraday effect is the induced magnetization of a medium ($\vec{M}_{ind}$) during circularly polarized excitation (green line). The direction of induced magnetization depends on the helicity of the light. In our experiments the optical rotation of a probe beam (purple line) indicates $\vec{M}_{ind}$. $\vec{B}_{app}$, external applied magnetic field; $\Delta\theta$, Faraday rotation angle; R(L)HCP, right- (left-) handed circularly polarized light.

In order to measure induced magnetization in plasmonic nanoparticles produced by the IFE, we performed static and ultrafast pump-probe Faraday rotation measurements on 100 nm diameter Au nanoparticle colloids (AuNP). First, the spectrally-resolved Verdet constant was determined from static Faraday rotation measurements on the sample colloid solution in a 1 cm-pathlength cuvette (Fig. 1a). On the same sample in 2 mm-pathlength flowing cell (Fig. 1b), the IFE was induced by a circularly polarized pump beam. The resultant magnetization was indicated by the Faraday rotation angle ($\Delta\theta$) of a linearly polarized probe beam. Based on the previously measured Verdet constant, the magnitude of the optical rotation of the probe thus enabled quantitative determination of the strength of the induced magnetization.

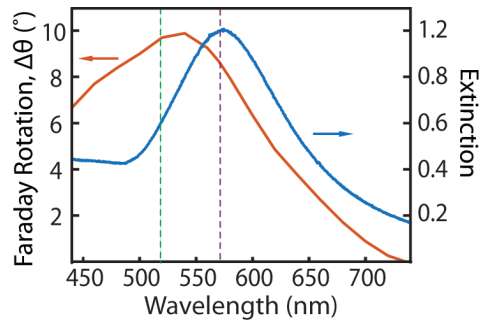

**Figure 2 | Static Faraday rotation of 100 nm diameter Au nanoparticle colloids.** Extinction spectra (blue) and Faraday rotation spectra (red). Dashed lines indicate pump (green) and probe (purple) wavelength used in the time-resolved pump-probe experiment.

In Fig. 2, the spectral dependence of the magnitude of Faraday rotation reflects the contribution of a plasmon resonance feature also observed in the extinction spectra, near 575 nm. Deviations between the extinction and the Faraday rotation result from the spectral overlap of the Au 5d-6s interband transition, which has also been observed, for example, from samples of Au-coated iron oxide nanocrystals[18]. As reported by several others[17,26,27], this trend indicates that Faraday rotation is strongly enhanced by the plasmon resonance, implying that the IFE should also be enhanced in these samples. The Verdet constant extracted at the concentration of AuNP in this study is 43.3 rad·T$^{-1}$m$^{-1}$ at 575 nm, which is about half that reported for the strong Faraday rotator materials such as terbium-doped boron-silicate glasses[28], and it is 16 times larger than bulk Au per unit length of metal[29].

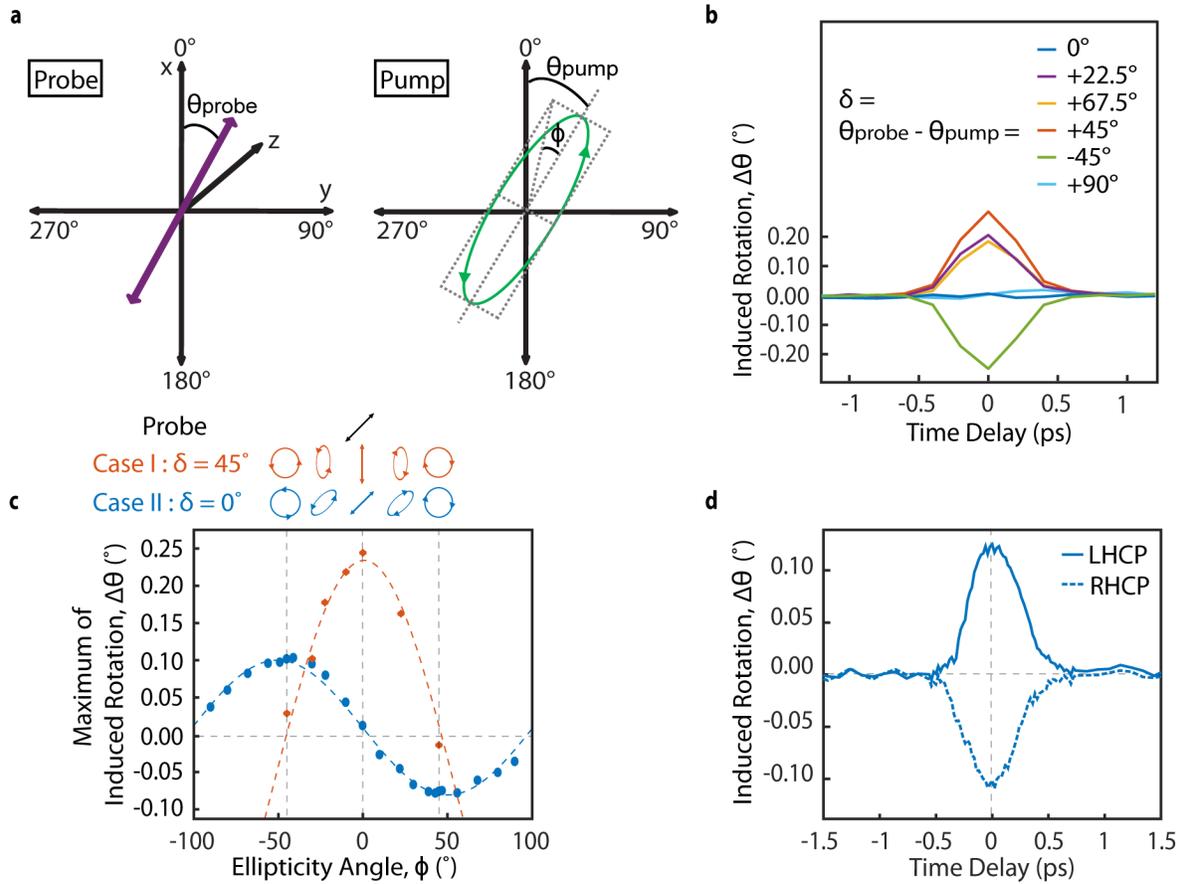

**Figure 3 | Time-resolved pump-probe IFE experiment on 100 nm diameter Au nanoparticle colloids. a**, Definition of the coordinate system and corresponding variables in equation (4). Light propagation is along the z-axis in the positive direction. The probe beam is linearly polarized for all experiments, with polarization angle $\theta_{probe}$. The pump beam can be linearly, elliptically, or circularly polarized. The polarization angle of the pump beam (when linearly polarized) or azimuth angle (when elliptically polarized) is $\theta_{pump}$. The difference between $\theta_{probe}$ and $\theta_{pump}$ is defined as $\delta = \theta_{probe} - \theta_{pump}$. The ellipticity angle, $\phi$ describes the extent of circular polarization of the pump and takes positive values when right-handed and negative values when left-handed. **b,** Light-induced rotation due to only the OKE ($\phi = 0°$) as a function of $\delta$. **c,** Light-induced rotation as a function of $\phi$. Red dots represent data from experiments with the geometry of Case I, in which $\delta = 45°$. Blue dots represent data from the Case II geometry, in which $\delta = 0°$. Dashed lines are the fit to the data. **d,** Light-induced rotation due to the pure IFE when the pump is circularly polarized with left-handed helicity ($\phi = -45°$, solid) or right-handed helicity ($\phi = +45°$, dashed), corresponding to the positive and negative maxima for Case II in **c**. A small OKE background signal at the same quarter wave plate rotation is subtracted. We believe that minor deviations from a pure IFE signal in the Case II geometry is due to imperfect alignment of the pump and probe beam along the sample axis.

An ultrafast pump-probe Faraday rotation measurement was made to quantify the induced magnetic field via the IFE (Fig. 1b). In this detection configuration, besides the optical rotation induced by the IFE, the optical Kerr effect (OKE) also caused rotation of the linearly polarized probe beam. Thus additional experiments were performed to extract the pure IFE signal. Similar pump-probe experiments employing a reflection geometry on bulk metal films have distinguished the signals originating from the OKE and IFE by analyzing the dependence of the optical rotation on the pump ellipticity[30]. The total induced rotation, $\Delta\theta$, from the combined effects of the IFE and the OKE depend on $\chi_{xxyy}$ and $\chi_{xyyx}$ of the third-order susceptibility as given by[31]:

$$\Delta\theta = [\text{Im}(F_+)\cdot\cos(2\phi)\cdot\sin(2\delta) + \text{Re}(F_-)\cdot\sin(2\phi)] \qquad (4)$$

with the contributions from the IFE and the OKE labeled as $F_-$ and $F_+$, and

$$F_\pm = \frac{-32\pi^2 L I_P}{\lambda c |1+n|^2}\left(\frac{\chi_{xxyy}\pm\chi_{xyyx}}{n}\right) \qquad (5)$$

Where $\phi$ is ellipticity angle of the pump, $\delta$ is the difference between the probe beam polarization angle and the pump azimuth angle, L is the effective travel distance of light inside the medium, $I_P$ is the pump intensity, $\lambda$ is the pump wavelength, and n is the refractive index.

The separate contributions of the OKE and IFE in the pump-probe Faraday rotation signal were disentangled by examining the dependence on $\delta$ and $\phi$ respectively. In the case with $\phi = 0°$ and $\delta$ varied, the measurement clearly showed that the signal at zero pump-probe time delay was correlated with $\delta$ (Fig. 3b). In agreement with equation (4) when $\phi = 0°$, this signal was maximized when $\delta = 45°$ and minimized when $\delta = 0°$ or 90°. Because we see $\sin(2\delta)$ dependence that is typical for the OKE (in contrast to four-fold sinusoidal dependence that can indicate a nonlinear anisotropic polarization effect)[32], we interpret this trend as being due entirely to the OKE. Based on this study, we assign two polarization configurations of the pump and probe beams as Case I in which the pure OKE signal is maximized with $\delta = 45°$ and Case II in which the pure OKE signal is absent with $\delta = 0°$. In the following discussion, the polarization configuration of the probe beam was fixed at 45°, and the ellipticity angle $\phi$ was varied while $\delta$ was maintained at 45° or 0° based on these two limiting cases.

Elliptical polarization was introduced to the pump and the results are summarized in Fig. 3c. In Case I ($\delta = 45°$), based on equation (4), $\Delta\theta = [\text{Im}(F_+)\cdot\cos(2\phi) + \text{Re}(F_-)\cdot\sin(2\phi)]$ and this matches our observation (red dots in Fig. 3c), for a situation with an OKE signal larger than an IFE signal. In our study the pure OKE is about ~2x larger than the pure IFE signal. In Case II ($\delta = 0°$), the OKE contribution is eliminated, and $\Delta\theta = \text{Re}(F_-)\cdot \sin(2\phi)$, indicating that the induced rotation is purely the result of the IFE (blue dots in Fig. 3c). With increasing ellipticity of the pump, the signal increases and a positive and negative maximum are observed when the pump is either purely left-hand or right-hand circularly polarized. This dependence on the helicity of the pump is an important criterion for confirming the presence of the IFE, because it indicates the direction of the induced magnetization that causes the optical rotation of the probe beam, consistent with equation (3). Further, the ellipticity-dependent studies (Fig. 3c) demonstrate a very clear difference between the signal from Case I experiments and Case II experiments.

The time response of the IFE with a small OKE background signal at the same quarter wave plate rotation subtracted is shown in Fig. 3d. We observe a single peak for any ellipticity of the pump, and this is in contrast with reports of Au thin films that show a bipolar shape of the time response at some elliptical pump excitations[30,33]. In addition, the time response appears to be limited by the pump-probe cross correlation, indicating an instantaneous magnetization and demagnetization process within the sub-picosecond time resolution of our experiment. While similar ultrafast demagnetization behavior has also been observed in other metal thin films[24,30,34], this behavior is in stark contrast to the optically-induced dynamic magnetism in ferromagnetic metal films and magnetic nanocrystals[24,35], where ultrafast demagnetization is followed by much slower hysteretic remagnetization. The time response observed here is consistent with a mechanism of coherent transfer of angular momentum from the optical field to the electronic motion, in agreement with the *ab initio* calculations discussed above[23].

In order to further confirm a magnetic field was created during optical excitation, as opposed to other non-linear optical phenomena or photothermal heating that could contribute to the measured optical rotation, we also performed the IFE experiment with counter-propagating pump and probe beams. In the polarization configuration of Case II (Fig. 4a), the induced rotation showed opposite sign when the pump beam was counter-propagated (red dots). The slightly smaller magnitude of rotation is due to a smaller effective increase of the pump-probe cross correlation in the counter-propagating configuration. However, in the configuration of Case I (Fig. 4b), the induced rotation does not depend on the pump beam direction. This result confirms that a transient change in refractive index of the material (OKE) or any other photothermal effects are reciprocal for either direction of the probe beam with respect to the pump beam. Importantly, in contrast, the IFE depends strongly on the propagation direction of the pump and probe beams, confirming the presence of the magnetization that gives rise to the nonreciprocal Faraday rotation of the probe beam.

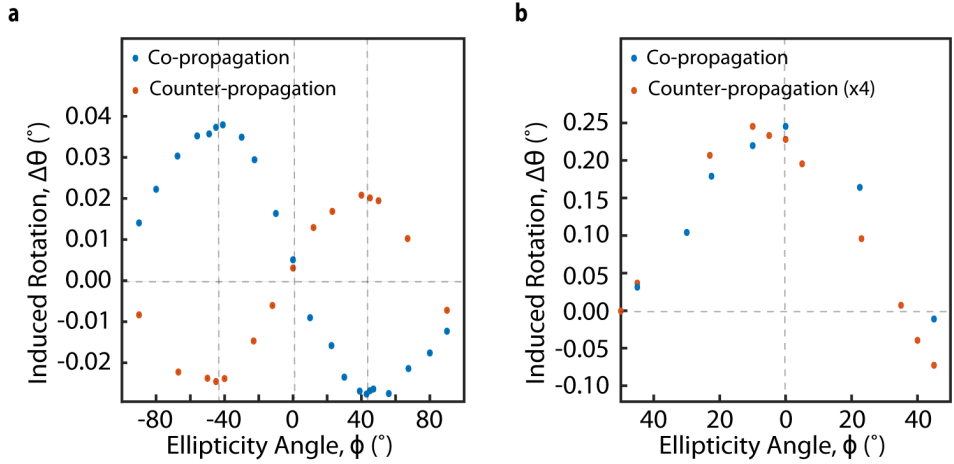

**Figure 4 | IFE Experiment with Counter-Propagating Pump and Probe Beams. a,** Case II configuration**.** The induced rotation has opposite sign when the pump beam is counter-propagated. This result demonstrates nonreciprocal Faraday rotation of the probe beam, resulting from the optically-pumped magnetization of the AuNP colloid due to the IFE. **b,** Case I configuration. The induced rotation resulting from the OKE does not depend on the propagation direction of the pump and probe beams.

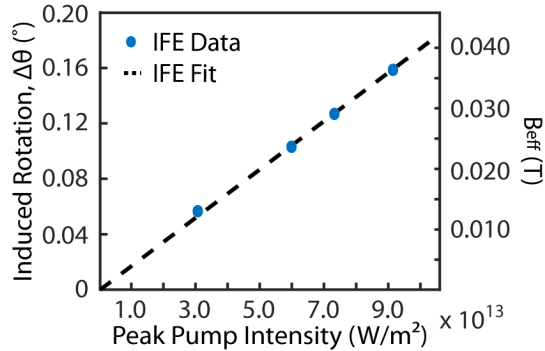

**Figure 5 | Dependence of Optical Rotation on Pump Intensity (Case II) and Corresponding Effective Magnetic Field. a,** Optical rotation as a function of peak pump intensity, with the polarization configuration of Case II. The fit shows a linear trend. The optical rotation is the result of an effective light-induced magnetic field, $\vec{B}_{eff}$, in the AuNPs.

The maximum induced optical rotation in the Case II configuration as a function of peak pump intensity is plotted in Fig. 5. The optical rotation shows a linear dependence, as predicted in equation (3) and equation (4). These observations provide further confirmation that the optical rotation of the probe beam is due to the optically-induced magnetization. The magnitude of induced magnetization, $\vec{M}_{ind}$, of the AuNPs during excitation can be determined by analyzing the measured Verdet constant and the induced optical rotation angle in the pump-probe experiments (see the SI for the full derivation).

$$\Delta\theta = \upsilon_{IFE}\vec{B}_{eff}L = \upsilon_{IFE}\mu_0\left(\frac{1}{\chi_v}+1\right)\vec{M}_{ind}L \tag{5}$$

Here $\vec{B}_{eff}$ can be thought of as the strength of an externally applied magnetic field that would be required to induce the same optical rotation observed in the pump-probe measurement[12]. We observe $\vec{B}_{eff}$ up to 0.036 T at a peak pump intensity of 9.1 x 10$^{13}$ W/m$^2$. The value of $\vec{M}_{ind}$ can be further analyzed to determine the magnetic moment of each Au nanoparticle. If we consider that the individual Au nanoparticles act as small bar magnets with rotation and diffusion that is much slower than the pulse duration, the ensemble response corresponds to an induced magnetic moment per particle of 2.6 x 10$^7$ $\mu_B$ or 8.3 x 10$^{-1}$ $\mu_B$ per Au atom, where $\mu_B$ is Bohr magneton. This induced magnetic moment is approximately one order of magnitude larger than the magnetic moment in other magnetic nanoparticles such as magnetite or $CoFe_2O_4$ colloidal nanoparticles[6,36]. Furthermore, the induced magnetic moment per Au atom in bulk Au films due to the IFE at the same incident optical power and wavelength is reported to be 5.9 x 10$^{-4}$ $\mu_B$, verifying the extraordinary >1,000-fold plasmonic enhancement of the optically induced magnetization measured here[15,30].

In summary, we report the first experimental observation of optically-induced magnetization in Au nanoparticles due to the IFE. We distinguished the contribution of the IFE and OKE in the optical rotation signal of a pump-probe experiment by analysis of the polarization-dependence, and by confirming the optical non-reciprocity imparted by the magnetization due to the IFE. Furthermore, the time response indicates a distinct mechanism of photo-induced magnetism that results from the coherent circular motion of electrons, in contrast with the spin dynamics typical in ultrafast studies of ferromagnetic materials. Additionally, we observed optically induced magnetization that is ~1,000 times larger than in bulk Au due to plasmonic field enhancement. We anticipate these results may be of great interest in the photonics community for application in ultrafast optical control of magnetic properties, and for all-optical methods of optical isolation that do not require externally applied magnetic fields.

**Acknowledgement**

The authors acknowledge the technical support of Dr. D. Rossi. This work is funded in part by the Gordon and Betty Moore Foundation through Grant GBMF6882 and by the Air Force Office of Scientific Research under award number FA9550-16-1-0154. M.S. also acknowledges support from the Welch Foundation (A-1886). D.H.S. appreciates the support from the Institute for Basic Science (IBS-R026-D1).

**Author contributions**

O.H.-C.C. carried out the measurement and analyzed the data. D.H.S. and M.T.S. supervised the project and participated in the analysis of the data.

**Competing interests**

The authors declare no competing interests.

## Methods

**Static Faraday rotation spectroscopy.** Experiments were performed on colloids of 100 nm diameter Au nanoparticles in water (OD = 1 at 572 nm, volume fraction = $1.14 \times 10^{-6}$, Alfa Aesar). The sample was placed in a 1 cm path length quartz cuvette. A static magnetic field (0.35 Tesla) was applied along the direction of a transmitted linearly polarized probe beam. A tungsten lamp was used as a light source, and the wavelength was selected by a monochromator (Newport, Model 74004). The beam was polarized by a sheet polarizer. A reference beam split by a reflective neutral density filter (OD = 0.2) was measured by a Si photodetector (Thorlabs, PDA100A). The Faraday rotation angle, $\Delta\theta$, was measured by sending the probe beam through Wollaston prism optically in-series with a balanced photodiode (New Focus, model 2307) after the beam had passed through the sample, and $\Delta\theta$ was calculated by the following equation: $\Delta I_{BPD}/I_{REF} = \sin(2\Delta\theta)$, where $\Delta I_{BPD}$ is power difference received at the balanced photodiode from the two polarized beams delivered by the Wollaston prism, $I_{REF}$ is power measured by a referenced Si photodetector. Note that $\Delta I_{BPD}/I_{REF} = \sin(2\Delta\theta) \approx 2\Delta\theta$ when $\Delta\theta$ is small (See Supplementary Information for detailed setup illustration).

**Time-resolved pump-probe Faraday rotation spectroscopy.** Measurements were made using an amplified Ti:sapphire laser system (KM Labs) operating at a repetition rate of 3kHz. The pump beam centered at $\lambda$=515 nm (green dashed line in Fig. 2) was generated from a home-built double-passed noncollinear optical parametric amplifier system (NOPA). The probe beam was derived from a white light continuum generated in a translating $CaF_2$ crystal and the wavelength was selected using a chirp compensating prism pair and a spatial filter. The beam size (FWHM) of the pump and probe beam, which were measured by a CCD camera (The Imaging Source, DMK 21BU618), are 150±6 μm and 80±2 μm in diameter, respectively. The temporal resolution of the measurements was 0.53 ps, which was determined by FWHM of the OKE signal of Au nanoparticle colloid (Supplementary Figure S2). The 100 nm diameter Au nanoparticle colloid was continuously circulated in a 2 mm path length quartz flow cell (Starna Cells, 45-Q-2) to prevent potential sample damage or accumulated thermal effects, as well as to reduce the effect of chirp in the pump beam. The induced Faraday rotation of the probe beam by the pump beam was measured using the same configuration as in the static Faraday rotation experiment as a function of pump-probe time delay, but without an externally applied magnetic field. The induced Faraday rotation angle was calibrated by using the combination of a neutral density filter and a linear polarizer (See Supplementary Information).

**Data Availability**

The data that support the plots within this paper and other finding of this study are available from the corresponding author upon reasonable request.

Supplementary Information for
1,000-Fold Enhancement of Light-Induced Magnetism in Plasmonic Au Nanoparticles

Oscar Hsu-Cheng Cheng[1], Dong Hee Son[1,2]*, Matthew Sheldon[1,3]*

[1]Department of Chemistry, Texas A&M University, TX, USA
[2]Center for Nanomedicine, Institute for Basic Science, Seoul, Republic of Korea
[3]Department of Material Science & Engineering, Texas A&M University, TX, USA
*e-mail: dhson@chem.tamu.edu; sheldonm@tamu.edu

## 1. Experimental setup

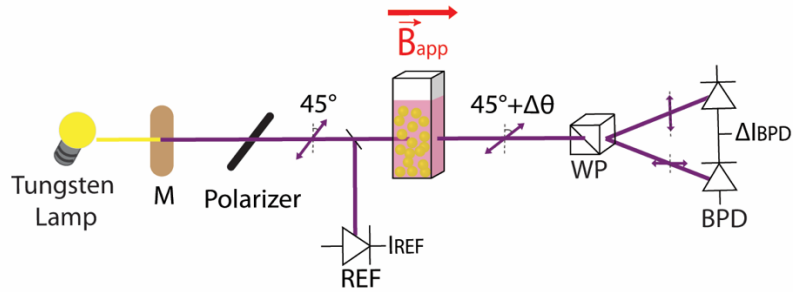

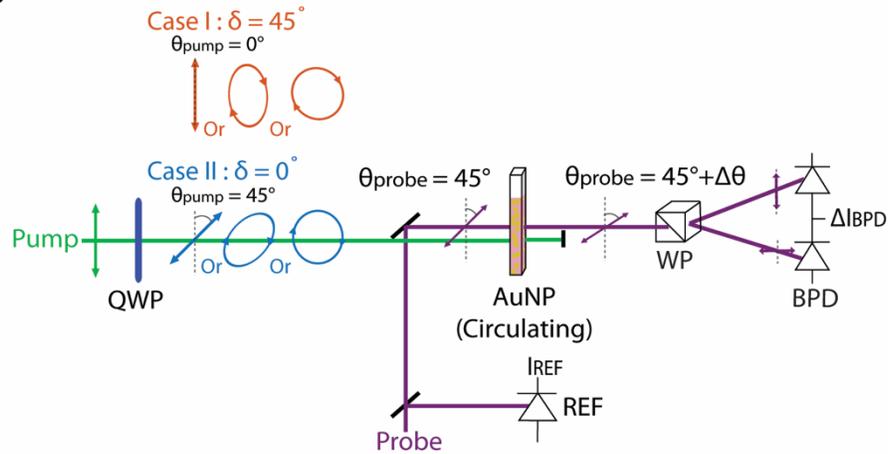

**Supplementary Figure S1: Detailed schematic illustration of experiment setup. a, Static Faraday rotation spectroscopy.** M, Monochromator; REF: Referenced Silicon photodiode; WP, Wollaston prism; BPD, Balanced photodiode. **b, Time-resolved pump-probe Faraday rotation spectroscopy.** The induced Faraday rotation was calibrated using the combination of a neutral density filter (ND = 0.06) and a linear polarizer. First, the neutral density filter was placed in front of one side of the BPD, resulting in some $\Delta I_{BPD}$ output signal. Then a linear polarizer was placed between AuNP and WP and was rotated until the BPD was balanced. $\Delta I_{BPD}/I_{REF}$ signal is then calibrated by the $\Delta I_{BPD}$ output signal produced by the neutral density filter and linear polarizer rotation angle. QWP, Quarter wave plate.

## 2. Cross-correlation of pump and probe beam

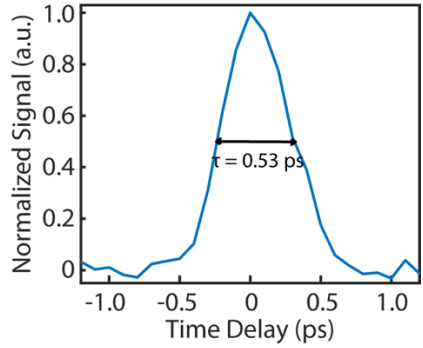

**Supplementary Figure S2: Estimation of the pump pulse width from the OKE signal of the Au nanoparticle colloids.** The OKE signal is treated as the cross-correlation of the pump and probe pulse, with full width at half maximum $\tau = 0.53$ ps. Thus we estimate the upper limit of pump pulse width to be $\tau/1.44 = 0.37$ ps, which is used to calculate peak pump intensity in Fig. 4.

## 3. Verdet constant of Faraday effect and the IFE in absorbing medium

The IFE expression in equation (1) does not account for the effect of dissipation and resonance in an absorbing medium. To estimate the effect of dissipation and resonance on the Verdet constant of Faraday effect and the IFE in this study, we consider the damping constant and resonance condition under a classical treatment[1]. The complex Verdet constant for gold colloid can be written as

$$\upsilon_{FE} \approx -\frac{\pi N e^3}{m_e^2 \lambda_v n_0} \frac{\omega}{\left(\omega_0^2 + i\gamma\omega - \omega^2\right)^2} = -5.15 \times 10^{-12} + 1.45 \times 10^{-11} i$$

where N is electron density of gold, $N = 5.57 \times 10^{28}$ m$^{-3}$, e is electron charge, $e = 1.602 \times 10^{-19}$ c, $m_e$ is effective mass of gold, $m_e = 9.02 \times 10^{31}$ kg, $\lambda_v$ is the wavelength of the light in vacuum, $\lambda_v = 5.15 \times 10^{-7}$ m, $n_0$ is the refractive index of the gold at 515 nm, $n_0 = 0.71912 + 2.0225i$, $\omega$ is frequency of the incident light, $\omega = 5.82 \times 10^{14}$ Hz, $\omega_0$ is the bulk plasma frequency, $\omega_0 = 1.32 \times 10^{16}$ Hz, and $\gamma$ is the bulk damping constant of gold, $\gamma = 1.26 \times 10^{14}$ Hz.

Then, the Verdet constant of the IFE can be written as

$$\upsilon_{IFE} = -\frac{\pi N e^3}{m^2 \lambda_v n_0} \frac{\omega}{\left(\omega^2 - \omega_0^2\right)^2 + \gamma^2 \omega^2} = -5.16 \times 10^{-12} + 1.45 \times 10^{-11} i$$

Thus the difference between $\upsilon_{FE}$ and $\upsilon_{IFE}$ is only 0.24% and is negligible in this study.

## 4. Calculation of the induced magnetization during the IFE experiment

In a Faraday rotation experiment, the applied magnetic field is related to the magnetization of the AuNP colloid by the following expression:

$$\vec{B}_{app} = \mu_0(\vec{H}+\vec{M}) = \mu_0\left(\frac{\vec{M}}{\chi_v}+\vec{M}\right) = \mu_0\left(\frac{1}{\chi_v}+1\right)\vec{M} \tag{S1}$$

where $\mu_0$ is permeability of vacuum, $\mu_0$ = 1.25664 × 10$^{-6}$ and $\chi_v$ is volume magnetic susceptibility of Au, $\chi_v$ = -3.44 × 10$^{-5}$. Therefore equation (2) in the main text can be written as:

$$\Delta\theta = \upsilon\vec{B}_{app}L = \upsilon_{FE}\mu_0\left(\frac{1}{\chi_v}+1\right)\vec{M}_{FE}L \tag{S2}$$

The measured magnetization of the AuNP colloid in the Faraday rotation experiment was -9.58 A/m. Since the Verdet constant, $\upsilon$, in the Faraday effect measurement and the IFE measurement is the same at the wavelength used in this study, we can obtain the induced magnetization during the IFE experiment by measuring the induced rotation $\Delta\theta$.

However, the attenuation of the pump intensity is not negligible, so the induced magnetization decreases along the optical path in the colloid. Based on equation (3) in the main text, the induced magnetization is linear with the pump intensity, so the measured overall Faraday rotation will be determined by the average pump intensity along the beam path, as shown below.

Based on the extinction coefficient $\varepsilon$ = 8.89 × 10$^{10}$ M$^{-1}$cm$^{-1}$ at the excitation wavelength 515 nm, the AuNP concentration was 6.37 × 10$^{-12}$ M. The excitation intensity, I, as a function of position, x, can be written as:

$$I(x) = I_0 10^{-A} = I_0 10^{-\varepsilon cx}$$

Since the effective magnetic field is linear with excitation intensity, we can write

$$\vec{M}_{ind}(x) = \vec{M}_{ind}^0 10^{-\varepsilon cx}$$

Then the induced rotation $\Delta\theta$ can be written as:

$$\Delta\theta = \int_0^{0.2} \upsilon_{IFE}\mu_0\left(\frac{1}{\chi_v}+1\right)\vec{M}_{ind}(x)dx$$
$$= \upsilon_{IFE}\mu_0\left(\frac{1}{\chi_v}+1\right)\int_0^{0.2} \vec{M}_{ind}^0 10^{-\varepsilon cx}\,dx$$
$$= 0.176\,\upsilon_{IFE}\mu_0\left(\frac{1}{\chi_v}+1\right)\vec{M}_{ind}^0$$

With 9.1 x 10$^{13}$ W/m$^2$ excitation intensity, and the induced rotation Δθ = 0.1587°, therefore $\vec{M}_{ind}^0$ = -0.98 A/m. This light-induced magnetization can be thought of equivalently in terms of what applied external magnetic field, $\vec{B}_{eff}$ in the main text, would result in the same magnetization, as commonly discussed in several other reports of the IFE[2,3]. Here, the maximum measured $\vec{B}_{eff}$ = 0.036T.

### 5. Calculation of magnetic moment per gold atom

With $\vec{M}_{ind}$ measured in the IFE experiment, the magnetic moment per gold nanoparticle $m_{AuNP}$ and magnetic moment per gold atom $m_{Au}$ can be calculated by

$$\vec{M}_{ind} = \frac{N_{AuNP} \cdot m_{AuNP}}{V} = \frac{N \cdot m_{Au}}{V \cdot N_{atom}}$$

where $N_{AuNP}$ is number of Au nanoparticles within the pump beam volume V, and $N_{atom}$ is the number of Au atoms per nanoparticle, $N_{atom} = 3.1 \times 10^7$.

### 6. Comparison of induced magnetic moment with other literature

The induced magnetic moment in Au films and AuNPs is compared, and the enhancement factor in the AuNPs is estimated. The Au film data is based on previous reports of pump-probe experiments on Au films. The calculations here for $m_{Au}$ are the same as in sections #4 and #5 above.

|  | Photon energy (eV) | Path length (cm) | Rotation (deg) | Intensity (x 10$^{13}$ W/m$^2$) | [b,c] $\vec{M}_{ind}$ (A/m) | [d] $m_{Au}(\mu_B)$ | Enhancement factor |
|---|---|---|---|---|---|---|---|
| V. V. Kruglyak et al.[4] | 1.55 | [a]1.3 x 10$^{-6}$ | 1 x 10$^{-3}$ | 12.6 | -128.9 | -2.36 x 10$^{-4}$ | 3800 |
| V. V. Kruglyak et al.[5] | 1.55 | [a]1.3 x 10$^{-6}$ | 2.5 x 10$^{-3}$ | 12.6 | -322.3 | -5.90 x 10$^{-4}$ | 1520 |
| This study | 2.41 | 0.2 | 1.59 x 10$^{-2}$ | 9.09 | -0.986 | -8.96 x 10$^{-1}$ | 1 |

[a] The path length in a gold film is estimated by the skin depth = (1/absorption coefficient).
[b] The Verdet constant of gold film is extracted from the report by J. McGroddy et al.[6]
[c] The induced magnetization is normalized to the same excitation intensity.
[d] The induced magnetic moment is normalized to the same excitation frequency, as shown in equation (3) in the manuscript.